\documentclass[aps,prd,twocolumn,groupedaddress,nofootinbib,showpacs,eqsecnum,showkeys,amsmath,amssymb,useAms]{revtex4}

\usepackage{euscript,graphicx,amssymb,subfigure}

\usepackage{epsfig}
\usepackage{amsfonts}
\usepackage{amsmath}
\usepackage{amssymb}
\usepackage{euscript}
\usepackage{color}
\usepackage{subfigure}

\def\lsim{\mathrel{\rlap{\lower3.5pt\hbox{\hskip0.5pt$\sim$}}
    \raise0.5pt\hbox{$<$}}}                
\def\gsim{~\rlap{$>$}{\lower 1.0ex\hbox{$\sim$}}}

\def\zB{\mbox{$z_{\rm BAO}$}}
\def\zCMB{\mbox{$z_{\rm CMB}$}}
\def\Om{\mbox{$\Omega_{\rm m0}$}}
\def\weff{\mbox{$w_{\rm eff}$}}
\def\peff{\mbox{$p_{\rm eff}$}}

\begin{document}

\title{Constraints from the CMB temperature and other common
observational data-sets on variable dark energy density models}

\author{Philippe Jetzer$^1$\footnote{\tt jetzer@physik.uzh.ch} and Crescenzo Tortora$^1$\footnote{{\tt ctortora@physik.uzh.ch}}}

\affiliation{$^1$ Universit$\ddot{a}$t Z$\ddot{u}$rich, Institut
f$\ddot{u}$r Theoretische Physik, Winterthurerstrasse 190,
CH-8057, Z$\ddot{u}$rich, Switzerland}

\begin{abstract}
The thermodynamic and dynamical properties of a variable dark
energy model with density scaling as $\rho_{x} \propto (1+z)^{m}$, z being
the redshift, are discussed following the outline of Jetzer et al.
\cite{Jetzer+11}. This kind of models are proven to lead to the
creation/disruption of matter and radiation, which affect the
cosmic evolution of both matter and radiation components in the
Universe. In particular, we have concentrated on the
temperature-redshift relation of radiation, which has been
constrained using a very recent collection of cosmic microwave
background (CMB) temperature measurements up to $z \sim 3$. For
the first time, we have combined this observational probe with a
set of independent measurements (Supernovae Ia distance moduli,
CMB anisotropy, large-scale structure and observational data for
the Hubble parameter), which are commonly adopted to constrain
dark energy models. We find that, within the uncertainties, the
model is indistinguishable from a cosmological constant which does
not exchange any particles with other components. Anyway, while
temperature measurements and Supernovae Ia tend to predict
slightly decaying models, the contrary happens if CMB data are
included. Future observations, in particular measurements of
CMB temperature at large redshift, will allow to give firmer bounds
on the effective equation of state parameter \weff\ of this kind of dark
energy models.
\end{abstract}





\maketitle

\section{Introduction}

The current standard cosmology model relies on the existence of
two unknown dark components, the so called ``dark matter'' (DM)
and ``dark energy'' (DE), which amount to $\sim 25\%$ and $\sim
70\%$ of the total energy budget in the Universe, respectively.
According to several observations, the Universe is spatially flat
and in an accelerated phase of its expansion \cite{perl, reiss,
debernardis, spe07, caldwell}. DE, described as a cosmological
constant $\Lambda$ in its simplest form, is modelled by a fluid
with a negative pressure, which is a fundamental ingredient to
explain the actual accelerated expansion of the Universe.

Several models have been proposed to explain DE \cite{PR03, Pad03,
D+05, CTTC06, Caldwell02, PR88, RP88, SS00}. An alternative
consists to consider a phenomenological variable DE density with
continuous creation/disruption of photons \cite{lima+96, LA99,
lima+00, puy, Jetzer+11} or matter \cite{CW90, ma}. The DE might
decay/grow slowly in the course of the cosmic evolution and thus
provide the source/sink term for matter and radiation. Different
such models have been discussed and strong constraints come from
very accurate measurements of the cosmic microwave background
(CMB) radiation and other typical cosmological probe.

CMB radiation is the best evidence for an expanding Universe
starting from an initial high density state. Within the
Friedmann-Robertson-Walker (FRW) models of the Universe the
radiation, after decoupling, expands adiabatically and scales as
$(1+z)$, $z$ being the redshift \cite{Weinberg72}. If we assume
that each component is not conserved, contrarily to the standard
scenario, then depending on the decay mechanism of the DE, the
created photons could lead to distortions in the Planck spectrum
of the CMB, and change the evolution of its temperature. The
chance to appreciate the deviation from the standard temperature
evolution is given by the increasingly number of recent works
collecting observations of CMB temperature both at low
\cite{battistelli, luz09} and higher redshifts
\cite{Noterdaeme+11}.

Following the theoretical lines of \cite{lima+96, LA99, lima+00,
puy}, in Jetzer et al. \cite{Jetzer+11} we have discussed a
variable DE model $\Lambda (z) \propto (1+z)^{m}$ decaying into
photons and DM particles. In particular we studied thermodynamical
aspects in the case of a continuous photon creation, which implies
a modified temperature redshift relation for the CMB. We have
tested the predicted temperature evolution of radiation with some
recent data on the CMB at higher redshift from both
Sunyaev-Zel'dovich (SZ) effect and high-redshift QSO absorption
lines.

In this paper we will present the results obtained in
\cite{Jetzer+11} within a more detailed theoretical background, by
further discussing the main hypothesis which concern the energy
conservation and thermodynamical laws. Furthermore, we will test
our model with an updated collection of data, joining our sample
with the new temperature observations reported in
\cite{Noterdaeme+11}. For the first time we will combine to the
temperature measurements also the data from different kind of
observations, like distance moduli of Supernovae Ia, observations
of the CMB anisotropy and the large-scale structure, together with
observational Hubble parameter estimations. We will dedicate a
particular attention to the estimate of both the actual matter
density parameter, the DE parameter $m$ and the effective equation
of state parameter $w_{\rm eff}$. The impact of each observational
probe on these estimates is also investigated.

The paper is organised as follows. In \S \ref{sec:theory} we will
present the model, while the data sets and fitting procedure are
described in \S \ref{sec:data}. The results are discussed in \S
\ref{sec:results} and \S \ref{sec:conclusions} is devoted to a
discussion of the results and conclusions.

\section{Theoretical framework}\label{sec:theory}

\subsection{Friedmann equations}

We assume a cosmological framework  based on the
usual Robertson-Walker (RW) metric element \cite{Weinberg72}
and that the Universe contains
three different components: a) a matter (both baryons and DM)
fluid, with equation of state $p_{m}=0$ (since $p_{m} <<
\rho_{m}$), b) a generalised fluid with pressure $p_{\gamma} =
(\gamma - 1) \rho_{\gamma}$, where $\gamma$ is a free parameter,
which is set to $4/3$ for a properly said radiation fluid, and c)
a DE, $x$ component, with pressure $p_x$ and density $\rho_x$. The
equation of state for the $x$ component could assume a very general
expression, but we limit ourselves to consider the simple linear relation
$p_{x} = w_{x} \rho_{x}$. We will set any 'bare' cosmological
constant $\Lambda_0$ equal to 0 \cite{Jetzer+11}. With these
components we get for the Einstein field equations
\cite{Weinberg72}

\begin{equation}
8\pi G \rho_{tot} = 3\frac{\dot R^2}{R^2}+3 \frac{k}{R^2}~,
\label{eq:Fried_1}
\end{equation}
\begin{equation}
8\pi G p_{tot} = -2\frac{\ddot R}{R}- \frac{\dot
R^2}{R^2}-\frac{k}{R^2}~, \label{eq:Fried_2}
\end{equation}
where $p_{tot} = p_{\gamma} +p_x$ and $\rho_{tot} = \rho_{m} +
\rho_{\gamma} + \rho_x$ are the total pressure and density, $R$ is
the scale factor, $k=0, \, \pm 1$ is the curvature parameter and a
dot means time derivative.
Furthermore, we will assume that there is no curvature,
thus $k=0$ \cite{debernardis}.

In the following we will adopt $w_x=-1$, however since we are
assuming that the vacuum decays into radiation and massive
particles, the effective equation of state $w_{eff}$, which is the
measured quantity, can differ from $-1$ (see later for further
details).

\subsection{Cosmological conservation laws and density evolution}

Following Jetzer et al. \cite{Jetzer+11} (but see also
\cite{lima+96, lima+00}) we set here the energy conservation
equation for the different fluids. The fluid as whole, verifies
the Bianchi identity, which means that energy and momentum are
locally conserved. In formulae it is $\nabla_{\mu} T^{\mu \nu} =
0$, with $T^{\mu \nu}$ the stress-energy tensor
\begin{equation}
T^{\mu\nu} = (\rho_{tot}+p_{tot})u^{\mu}u^{\nu}-p_{tot}g^{\mu\nu}
\end{equation}
with $u^{\mu}$ being the four velocity. After easy calculations,
this conservation law reads
\begin{equation}
\dot\rho_{tot} +3 (\rho_{tot}+p_{tot}) H = 0  ~,\label{eq:cons_ALL}
\end{equation}
where $H = \dot R / R$ is the Hubble parameter. In the standard
approach, each component is conserved, thus Eq.
(\ref{eq:cons_ALL}) holds for all the fluid components, but here
we will suppose that each component will exchange energy with each
other. In particular, for matter, radiation and DE we impose the
following relations (see also \cite{mubasher})
\begin{equation}
\dot\rho_{m} +3 \rho_{m} H = (1-\epsilon) ~ C_{x}
~,\label{eq:cons_mat}
\end{equation}
\begin{equation}
\dot\rho_{\gamma} +3 \gamma \rho_{\gamma} H = \epsilon ~ C_{x}
~,\label{eq:cons_gamma}
\end{equation}
\begin{equation}
\dot\rho_{x} +3(p_{x} + \rho_{x})H = - C_{x} ~,\label{eq:cons_x}
\end{equation}
where we assume that both the matter and the radiation fluids
exchange energy with the DE as parameterized by $C_{x}$ and
$\epsilon$. In particular, $C_{x}$ depends on the DE and, indeed,
acts as a source/sink term for the fluids energy. Evidently, if no
interaction between the different components exists, then $C_x$ is
null and the standard picture is recovered. Moreover, if $\epsilon
= 0$ ($\epsilon = 1$), then the DE exchanges all the energy with
matter (radiation). $C_x$ can describe different physical
situations such as, for instance, a thermogravitational quantum
creation theory \cite{LA99} or a quintessence scalar field
cosmology \cite{RP88}. As we will discuss later on, $\epsilon$ has
to be very small (i.e., $\epsilon \ll 1$), otherwise the radiation
density would become much too big, contrary to present values. As
an order of magnitude estimate we expect $\epsilon \simeq
\frac{p_{\gamma} + \rho_{\gamma}}{\rho_{m}}$, for which indeed
$\epsilon \ll 1$, since $p_{\gamma} + \rho_{\gamma} \ll \rho_{m}$.

Adopting as mentioned above the relation $p_x=-\rho_x$ and defining
$\rho_x=\Lambda(t)/(8\pi G)$, from Eq.
(\ref{eq:cons_x}) we obtain
\begin{equation}
C_x= -\frac{\dot\Lambda(t)}{8\pi G}~. \label{eq:Cx}
\end{equation}
We assume a power law model for the $\Lambda(t)$ function,
$\Lambda(t) = B (R / R_{0})^{-m}$, or equivalently in terms of
redshift, $\Lambda(z) = B (1+z)^{m}$, where $B$ is a constant. The
value of this constant is $B=3 H_{0}^{2} (1- \Om)$, which can be
found using Eq. (\ref{eq:Fried_1}) at the present epoch, assuming
that today $\rho_{\gamma} \ll \rho_{m}, \rho_{x}$. Thus, the
density evolution for the $x$ component is given by
\begin{equation}
\rho_{x}/\rho_{crit}  = (1 - \Om) (1+z)^{m},
\end{equation}
where we have defined as $\rho_{crit} = \frac{3 H_{0}}{8 \pi G}$
the present critical density of the Universe. If $m$ is positive,
then the DE slowly decreases as a function of the cosmic time,
whereas if $m$ is negative the inverse process happens.

From Eqs. (\ref{eq:cons_mat}) and (\ref{eq:cons_gamma}) we derive
the evolution laws for matter and radiation,
\begin{align}\label{eq:rho_m}
\rho_{m}/\rho_{crit} & = \Om (1+z)^{3}
\\ \nonumber
& \quad - (1-\epsilon) \frac{m (1- \Om)}{m - 3} [(1+z)^{m} -
(1+z)^{3}]~,
\end{align}
\begin{align}\label{eq:rho_gamma}
\rho_{\gamma}/\rho_{crit} & = \Omega_{\gamma 0} (1+z)^{3 \gamma}
\\ \nonumber
& \quad - \epsilon \frac{m (1- \Om)}{m - 3 \gamma} [(1+z)^{m} -
(1+z)^{3 \gamma}]~,
\end{align}
where \Om\ and $\Omega_{\gamma 0}$ are the matter and radiation
energy densities at $z=0$, respectively.

Since we are interested in the evolution of the radiation
temperature, it is useful to discuss the extreme case when only
photons enter in the process (i.e. $\epsilon = 1$). Then
for the matter density in Eq. (\ref{eq:rho_m}) the usual evolution
$\propto (1+z)^{3}$ holds, while the radiation besides the usual
term $\Omega_{\gamma 0} (1+z)^{3 \gamma}$ has also a perturbative term
depending on $m$. Since today \footnote{The present radiation density
is the only cosmological parameter accurately measured. The
radiation density is dominated by the energy in the cosmic
microwave background (CMB), and the COBE satellite FIRAS
experiment determined its temperature to be $T = 2.725 \pm 0.001
\, K$ \cite{Mather+99}, corresponding to $\Omega_{\gamma 0} \sim  5
\times 10^{-5}$.} $\Omega_{\gamma 0}   \sim 5 \times 10^{-5}$
it turns out that $m$ has to be extremely small $\lsim 10^{-4}$.
Therefore, unless $m$ is extremely small or vanishing, DE
has to decay mainly in matter with possibly some photons as well.
Thus the condition $\epsilon \ll 1$ has to hold.

\subsection{Hubble and deceleration parameter}

Due to the very small value of $\Omega_{\gamma 0}$ it follows that
the evolution of the Universe is essentially driven by the DE and
DM, therefore, from Eq. (\ref{eq:Fried_1}) the following law for
the Hubble parameter holds
\begin{align}\label{eq:H}
H(z) & \simeq  \frac{8 \pi G}{3} (\rho_{m} + \rho_{x})
\\ \nonumber
& \hspace{-0.7cm} =  H_0 \left[\frac{3 (1-\Om)}{3 - m} (1+z)^{m} +
\frac{(3\Om - m) }{3-m} (1+z)^{3} \right]^{1/2},
\end{align}
which is obviously the same expression found in Ma \cite{ma}.

Recasting Eq. (\ref{eq:cons_x}), it is possible to write
\begin{equation}
\dot\rho_{x} +3H(p_{x} + \rho_{x} +\frac{C_{x}}{3 H}) = 0
~,\label{eq:cons_x_2}
\end{equation}
which shows that the term $C_x$ contributes to an effective
pressure
\begin{equation}
\peff = p_{x} + \frac{C_{x}}{3 H} = -\rho_{x} +\frac{C_{x}}{3 H}~.
\end{equation}
Therefore, we get an equivalent
effective DE equation of state $\weff$ \cite{ma}
\begin{equation}
\weff = \frac{\peff}{\rho_x} = \frac{m}{3}-1 .
\end{equation}
If $m>0$ then we have $\weff
> -1$, i.e. our model is quintessence-like \cite{PR88,RP88,D+05},
while we have a phantom-like \cite{Caldwell02} model when $m$ is
negative and $\weff<-1$. Another interesting quantity is the
deceleration parameter, which can be written as
\begin{align}
q(z) & = - \frac{\ddot{R}R}{\dot{R}^{2}}= \\ \nonumber &
\frac{(1+z)^{3}(m-3\Om)+3(m-2)(1+z)^{m}(\Om-1)}{2(1+z)^{3}(m-3\Om)+6(1+z)^{m}(\Om-1)}~.
\end{align}
Imposing that $q(z)=0$, we can determine the {\it transition
redshift}, i.e. the redshift at which the Universe changed from a
deceleration to an acceleration phase, which is given by
\begin{equation}
z_{T}=\bigg ( \frac{3(2-m)(1-\Om)}{3\Om-m}
\bigg)^{\frac{1}{3-m}}-1~.
\end{equation}
From this result we see that the larger $m$ is, the earlier the
Universe changes from deceleration to acceleration.

\begin{figure*}[t]
\epsfig{file=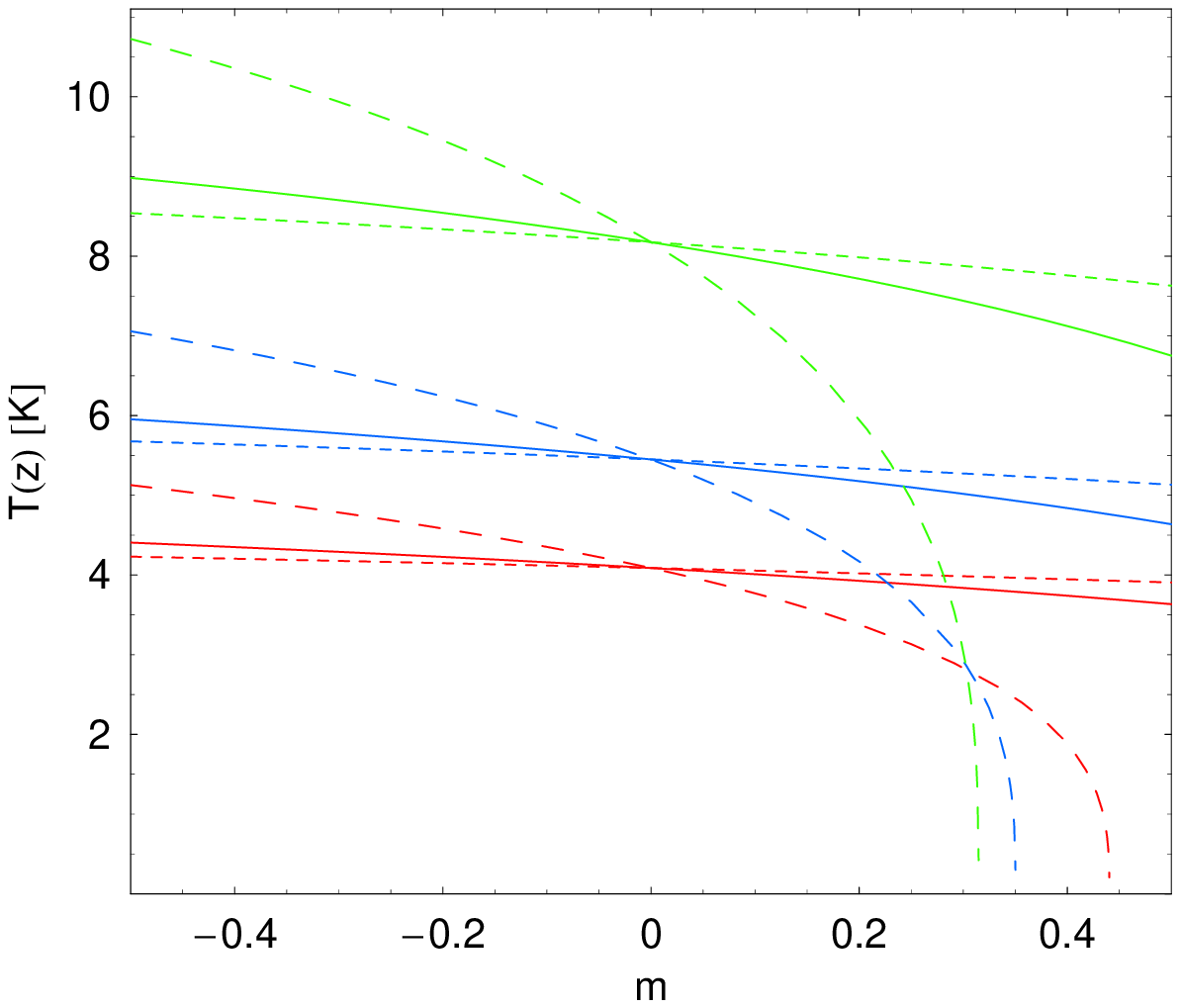, width=0.3\textwidth}
\epsfig{file=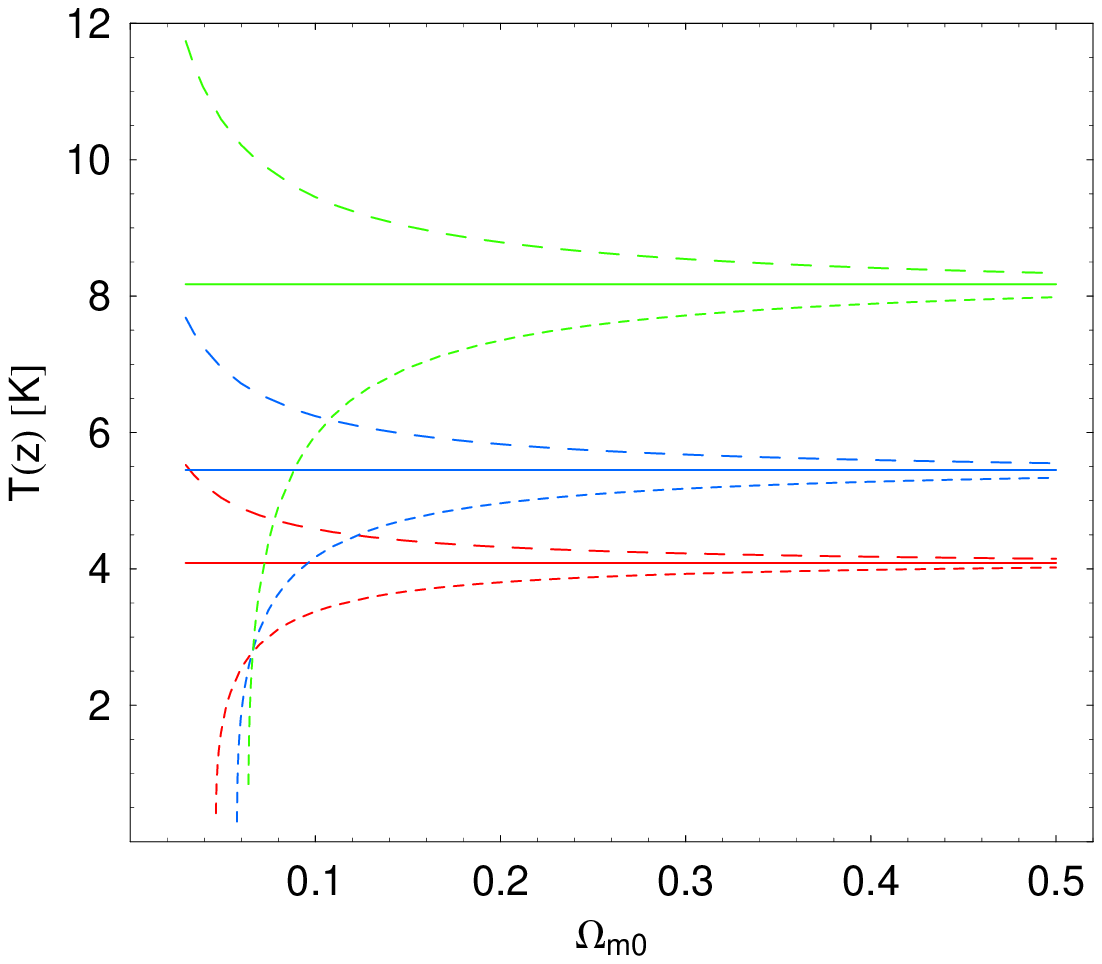, width=0.3\textwidth}
\epsfig{file=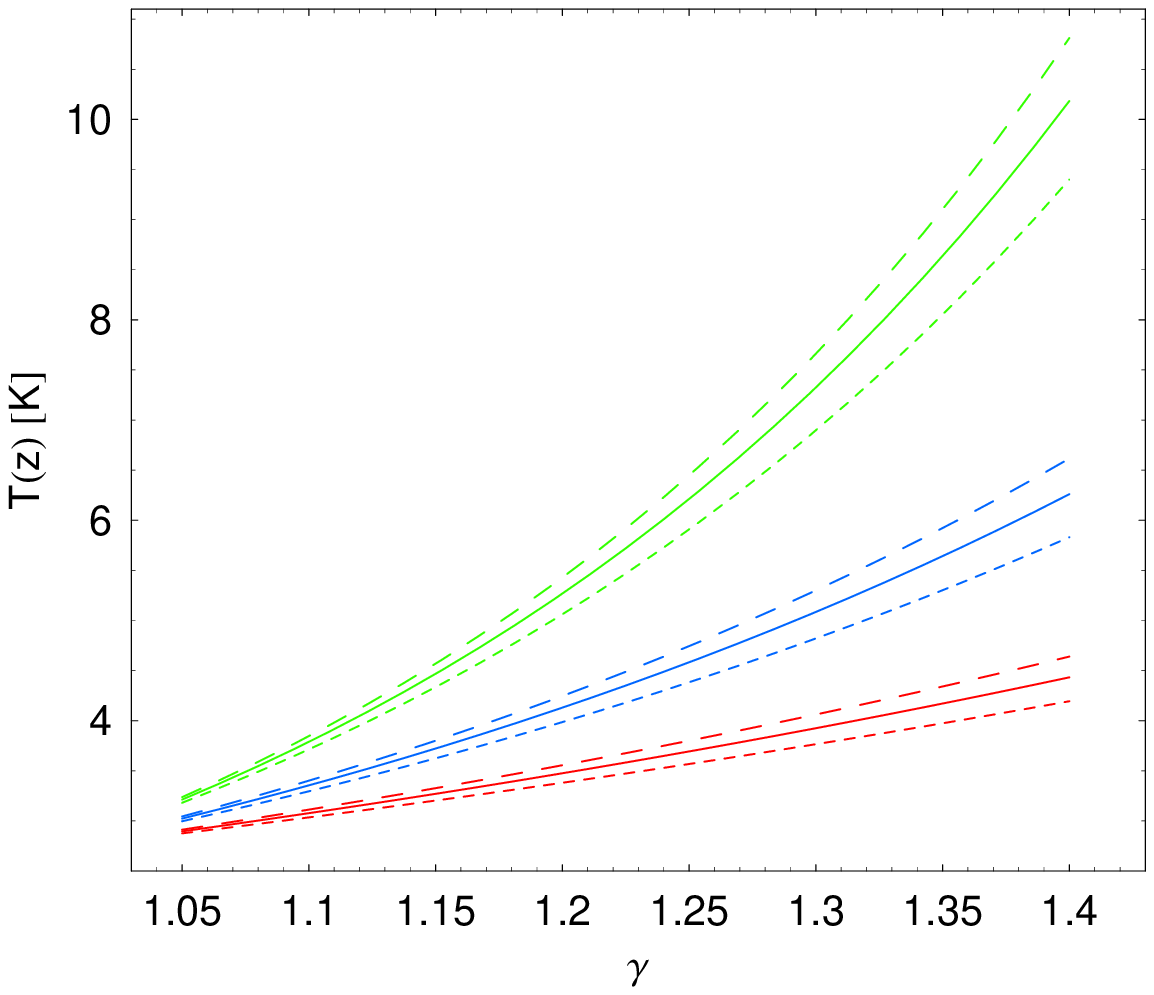, width=0.3\textwidth} \caption{Radiation
temperature in Eq. (\ref{ma3}) in terms of $m$, \Om\ and $\gamma$.
Red, blue and green lines are for $z=0.5$, $z=1$ and $z=2$,
respectively. {\it Left.} $T$ as a function of $m$. $\gamma = 4/3$
and $\Om = 0.1$ (long-dashed), $=0.3$ (continue) and $=0.5$
(short-dashed). {\it Middle.} $T$ as a function of \Om. $\gamma =
4/3$ and $m = -0.2$ (long-dashed), $=0$ (continue) and $=0.2$
(short-dashed). {\it Right.} $T$ as a function of $\gamma$.
$\Om=0.273$ and $m = -0.2$ (long-dashed), $=0$ (continue) and
$=0.2$ (short-dashed).} \label{fig:fig1}
\end{figure*}

\subsection{Thermodynamical aspects and CMB temperature evolution}\label{sec:T_relation}

We follow here the approach outlined in Lima \cite{lima+96}, where
he defines a current as $N^{\alpha}=n u^{\alpha}$ with $n$ being
the particle number density of the photons or of the DM particles.
Indeed, there is a current for each of these components. Due to
the decaying vacuum the current satisfies the following balance
equation (one for each component)
\begin{equation}
\dot n_i + 3n_iH= \psi_i~, \label{nn}
\end{equation}
with $i=\gamma$ or $DM$ ($\gamma$ for the photons and $DM$ for the
dark matter) and $\psi_i$ is the corresponding particle source.
For decaying vacuum models $\psi_{\gamma}+\psi_{DM}$ is positive
and related to the rate of change of $\rho_x$. We can also define
an entropy current of the form
\begin{equation}
S^{\alpha}=\sum_i n_i \sigma_i u^{\alpha}_i~,
\end{equation}
where $\sigma_i$ is the specific entropy per particle (photons, DM
and in principle also DE). If the DE $\rho_x$ is constant the
above entropy current is conserved. The existence of a non
equilibrium decay process of the vacuum implies
$S^{\alpha}_{;\alpha} \geq 0$, thus an increase of the entropy as
a consequence of the second law of thermodynamics. In principle
the second law should be applied to the system as a whole, thus
including the vacuum component \cite{mubasher}. Assuming that the
vacuum is like a condensate with zero chemical potential
$\mu_{vac}$ it follows from Euler's relation
\begin{equation}
\mu_{vac}= \frac{\rho_x+p_x}{n}-T \sigma_{vac}~,
\end{equation}
provided $w_x=-1$, that $\sigma_{vac}=0$ and thus
its contribution to the entropy current vanishes.
Given our assumptions the vacuum plays the role
of a condensate carrying no entropy.

For instance, for quintessence models in the limit where the
scalar field does not depend on time, and thus its time derivative
vanishes, one gets $w_x=-1$. In the later stages of the Universe
the time dependence is possibly very weak so that $w_x=-1$ holds
up to small corrections.

The equation for the particle number density of radiation
component is given by Eq. (\ref{nn}) with $i=\gamma$. Using Gibbs
law and well-known thermodynamic identities, following the
derivation given in the paper by Lima et al. \cite{lima+00}, one
gets (see also \cite{lima+96, Jetzer+11})
\begin{equation}
\frac{\dot T}{T} = \left(\frac{\partial p_{\gamma}}{\partial
\rho_{\gamma}}\right)_n \frac{\dot n_{\gamma}}{n_{\gamma}} -
\frac{\psi_{\gamma}}{n_{\gamma} T \left(\frac{\partial
\rho_{\gamma}}{\partial T}\right)_n}
\left[p_{\gamma}+\rho_{\gamma}- \frac{n_{\gamma} \epsilon \,
C_x}{\psi_{\gamma}}\right]~. \label{1}
\end{equation}

To get a black-body spectrum the second term in brackets in Eq.
(\ref{1}) has to vanish, thus

\begin{equation}
\epsilon \, C_x = \frac{\psi_{\gamma}}{n_{\gamma}} \left[
p_{\gamma}+\rho_{\gamma} \right]~. \label{3}
\end{equation}

Thus, Eq. (\ref{1}) becomes

\begin{equation}
\frac{\dot T}{T} = \left(\frac{\partial p_{\gamma}}{\partial
\rho{\gamma}}\right)_n \frac{\dot n_{\gamma}}{n_{\gamma}}~.
\label{5}
\end{equation}

With $\left(\frac{\partial p_{\gamma}}{\partial
\rho_{\gamma}}\right)_n =(\gamma -1)$ one obtains

\begin{equation}
\frac{\dot T}{T} = (\gamma -1) \frac{\dot
n_{\gamma}}{n_{\gamma}}~. \label{6}
\end{equation}

Using the equation for the particle number conservation Eq.
(\ref{nn}) into Eq. (\ref{6}) leads to

\begin{equation}
\frac{\dot T}{T} = (\gamma -1)
\left[\frac{\psi_{\gamma}}{n_{\gamma}}-3H\right]~. \label{eq:T_Psi}
\end{equation}

With Eqs. (\ref{3}) and (\ref{eq:T_Psi}) we get

\begin{equation}
\frac{\dot T}{T} = (\gamma -1) \left[-\frac{\epsilon \,
\dot\Lambda}{8\pi G(p_{\gamma}+\rho_{\gamma})}-3H\right]~.
\label{eq:T_Psi_2}
\end{equation}
Now, following the previous discussion on $\epsilon$ and
aiming to be very general, we set $\epsilon =
\frac{\rho_{\gamma}+p_{\gamma}}{\rho_{m}} \tilde{\epsilon}$, where
$\tilde{\epsilon}$ is a new parameter and insert it into Eq.
(\ref{eq:T_Psi_2}). Taking the sum of Eqs. (\ref{eq:Fried_1}) and
(\ref{eq:Fried_2}) we find
\begin{equation}
8\pi G(\rho_{tot}+p_{tot}) \simeq 8\pi G \rho_{m} =2\frac{\dot
R^2}{R^2}-2 \frac{\ddot R}{R} = -2\dot H~. \label{13}
\end{equation}
Finally, we obtain the expression
\begin{equation}
\frac{\dot T}{T} = (\gamma -1) \left[\frac{\dot\Lambda
\tilde{\epsilon}}{2\dot H}-3H\right]~, \label{14}
\end{equation}
which we can integrate

\begin{equation}
\int^{t_0}_{t_1} \frac{\dot T}{T} dt = (\gamma - 1)
\int^{t_0}_{t_1}\left[\frac{\dot\Lambda \tilde{\epsilon}}{2\dot
H}-3H\right] dt~, \label{15}
\end{equation}
where $t_0$ denotes the present time and $t_1$ some far instant in
the past. Indeed, if $\dot\Lambda$ vanishes and $\gamma=4/3$ one
gets the usual dependence $T(t)=\frac{R(t_1) T(t_1)}{R(t)}$ for a
radiation fluid. To carry out the integration of the first term on
the right hand side it is useful to perform a change of variable
from $t$ to $z$ and accordingly $\frac{dt}{dz}=\frac{-1}{H(1+z)}$.
This way we get (with $z_1$ corresponding to the time $t_1$ and
$z_0=0$ corresponding to $t_0$ present time)
\begin{align}\label{eq:T_fin}
ln \frac{T(z=0)}{T(z_1)}+3(\gamma-1)ln\frac{R(z=0)}{R(z_1)}=
\\ \nonumber \frac{(\gamma-1)}{2} \int_0^{z_1}
\frac{\Lambda^{\prime} \tilde{\epsilon}}{H^{\prime} H(1+z)} dz~,
\end{align}
where $^{\prime}$ denotes derivative with respect to $z$.

As next we insert $H(z)$ and its derivative as taken from Eq.
(\ref{eq:H}) into Eq. (\ref{eq:T_fin}) and integrate it, to get
(setting $z_1=z$)

\begin{equation}
T(z) = T_0 \left(\frac{R_0}{R(z)}\right)^{3(\gamma-1)}
exp\left(\frac{B(1-\gamma)\tilde{\epsilon}}{3H_0^2(\Om-1)}A\right)~,
\label{eq:T_fin_2}
\end{equation}
where
\begin{align}\label{ma2}
A = ln((m-3\Om)+m(1+z)^{m-3}(\Om-1))
\\ \nonumber  -ln((m-3)\Om) ]~.
\end{align}

We can also write Eq. (\ref{eq:T_fin_2}) as
\begin{align}\label{ma3}
T(z) & = T_0 (1+z)^{3(\gamma-1)} \\ \nonumber & \times
\left(\frac{(m-3\Om)+m(1+z)^{m-3}(\Om-1)}{(m-3)\Om}\right)^{\tilde{\epsilon}(\gamma-1)}~.
\end{align}
We inserted in the exponent of Eq. (\ref{eq:T_fin_2}) the explicit
form of $B$, thus getting as exponent in the above Eq.
$\tilde{\epsilon} (\gamma-1)$. Hereafter, we will set
$\tilde{\epsilon} = 1$. Clearly $\tilde{\epsilon}$ and $m$ are not
independent, we checked using the temperature redshift data that
if $\tilde{\epsilon}$ is bigger ($\sim 10$ or more), then $m$ has
to be extremely small consistently with what mentioned in section
II.B (as it would lead to a too high production of photons in the
DE decay). On the other hand, if $\tilde{\epsilon}$ gets smaller
(e.g., $\sim 0.1)$ $m$ gets bigger ($\sim 0.2$) and accordingly
$w_{eff}$, moreover $m$ would be poorly constrained, since the
uncertainties would then be very high. But from the other data
(without the the temperature ones) there are already stringent
limits on $m$ and thus this way one could get lower limits on
$\tilde{\epsilon}$, under the assumption that DE decays also in
photons.

Notice that for $z=0$ we have $T(0)=T_0$,
whereas for $m=0$ the expression in the parenthesis is equal to 1
and thus $T(z)=T_0(1+z)^{3(\gamma-1)}$, which for the canonical
value of $\gamma=4/3$ reduces to the standard expression.

In Fig. \ref{fig:fig1}, the temperature relation from Eq.
(\ref{ma3}) is discussed in terms of the model parameters and
redshift. At fixed redshift, the temperature is a decreasing
function of $m$, i.e. larger $m$ means a colder CMB radiation
temperature and when compared with the unperturbed case with
$m=0$, then a decaying/increasing model predicts colder/hotter
temperatures. The temperature is sensitive to \Om\ mainly at $\Om
\lsim 0.1-0.2$ and less at larger \Om, while it is a strong function
of $\gamma$, increasingly as a function of redshift.

\section{Fitting procedure and data}\label{sec:data}

To constrain the parameters of the model we use a set of different
kinds of measurements. Our data sets include the measurements of
CMB temperatures from high redshift quasars and SZ effect, high
quality ``UnionII'' SN Ia data, baryon acoustic oscillation
measurement from the Sloan Digital Sky Survey, the shift parameter
from \textrm{WMAP} three years results and 9 observational $H(z)$
data. To break the degeneracies between the parameters and explore
the power and differences of the constraints for these data sets,
we use them in several combinations to perform our fitting.

In order to constrain the model parameters we maximize the
likelihood function $\mathcal{L}({\bf p}) = \exp\left[-\frac{1}{2}
\chi^{2} \right]$, where ${\bf p}$ denotes the set of model
parameters and $\chi^{2}$ is a suitable merit
function\footnote{This is equivalent to a $\chi^2$ minimization.}.
The isolikelihood (or the iso$\chi^2$) contours provide
constraints on the parameter space. The $68\%$ confidence levels
(CL) are obtained by imposing $\Delta \chi^{2}=
\chi^{2}-\chi^{2}_{min} = 1$ and $2.3$ for $n_{p}= 1$, $2$ free
parameters, where $\chi^{2}_{min}$ is the minimum of $\chi^{2}$
function. The $90\%$ CL, is given by $\Delta \chi^{2} = 2.71$ and
$4.61$ for $n_{p}= 1$, $2$. Finally, the $95\%$ CL is given by
$\Delta \chi^{2} = 4$ and $6.17$ for $n_{p}= 1$, $2$. In order to
give a quantification of the errors on a given parameter we can
follow different approaches. When $n_{p} = 1$, then the error on
the parameter is determined simply adopting the above conditions
on $\Delta \chi^{2}$. On the contrary, if $n_{p} = 2$ (which is
one of the cases we will investigate in the paper), to constrain a
given parameter $p_{1}$, we rely on the marginalized function
defined as $\mathcal{L}_{p_1}(p_{1}) \propto \int_{p_{2}} dp_{2}
\mathcal{L}({\bf p})$, which is normalized to 1 at the maximum.
The value of $p_1$ corresponding to the maximum of such a function
is chosen as our best fitted value\footnote{Note that for
asymmetric CLs, the $\chi^{2}$ minimum and the maximum of the
marginalized likelihood can be different.} and the CLs are
determined by applying to this marginalized likelihood the
conditions above on $\Delta \chi^{2}$ for $n_{p} = 1$.

In the course of the paper, for sake of simplicity we will report
only the 68\% CL for the listed best fitted parameters.

\subsection{Temperature measurements (T)}

To test the temperature evolution for the radiation component, we
rely on the CMB temperatures derived from the absorption
lines of high redshift systems and the ones from SZ effect in
clusters of galaxies (we will collectively quote as $T_{CMB}$,
hereafter). At high redshift the CMB temperature is recovered from
the excitation of interstellar atomic or molecular species that
have transition energies in the sub-millimetre range and can be
excited by CMB photons. When the relative population of the
different energy levels are in radiative equilibrium with the CMB
radiation, the excitation temperature of the species equals that
of the black-body radiation at that redshift, providing one of the
best tools for determining the black-body temperature of the CMB
in the distant Universe \cite{ge97, sri00, mol02, puy93, gal98,
sta98, cui05, sri08}. In Jetzer et al. \cite{Jetzer+11}, we
adopted a sample of 5 QSO adsorption measurements, which is now
updated to 9 after the recent measurements reported in Noterdaeme et al.
\cite{Noterdaeme+11}. In summary we have 4 data points from the
analysis of the fine structure of atomic carbon (AC) and 5
measurements based on the rotational excitation of CO molecules
(CO) \cite{Noterdaeme+11}.

At lower redshift we use the measurements from the SZ effect.
During passage through a cluster of galaxies some of the photons
of the CMB radiation are scattered by electrons in the hot
intracluster medium. This imprint was first described by SZ
\cite{sun72}. Thus, spectral measurements of galaxy clusters at
different frequency bands yield independent intensity ratios for
each cluster. The combinations of these measured ratios permit to
extract the cosmic microwave background radiation (see Fabbri et
al. \cite{fab78}). We will rely on the data compilation in Luzzi
et al. \cite{luz09}, which have analyzed the results of
multifrequency SZ measurements toward several clusters from 5
telescopes (BIMA, OVRO, SUZI II, SCUBA and MITO).

We will match the observed $T_{CMB}$ with the theoretical
expression $T_{th}$, which we have derived in Eq. (\ref{ma3}), by
minimizing the following merit function
\begin{equation}
\chi^{2}_{TCMB}= \sum_{i=1}^{N_{TCMB}} \bigg (\frac{T_{th}^{i}-
T_{CMB}^{i}}{\sigma_{CMB,i}} \bigg)^{2}~,
\end{equation}
where $\sigma_{CMB,i}$ is the error on the temperature estimates
and $N_{TCMB} = 22$ is the number of available observational data.

\subsection{High quality Supernovae Ia data set
(SN)}\label{sec:SN}

The most important candle we use is the type Ia supernovae (SN).
We adopt the UnionII dataset discussed in Amanullah et al.
\cite{Amanullah+08}, which consists of $N_{SN} = 557$ datapoints
from $z=0$ to $z=1.4$, compiled after the combination of different
datasets and the consequent application of various selection cuts
to create a homogeneous and high signal-to-noise sample.

The data points for SN are given in terms of distance modulus
$\mu_{obs} = m - M$, where $m$ and $M$ are the apparent and
absolute magnitude, respectively. The theoretical distance modulus
is given by
\begin{equation}
\mu_{th} (z) = 5 \log_{10}D_{L}(z) + \mu_{0}~,
\end{equation}
where $\mu_{0} = 42.38 - 5 \log_{10}h$ and $D_{L}(z)$ is the
luminosity distance at the redshift $z$. The $\chi^{2}$ function
to be minimized is
\begin{equation}
\chi _{\mathrm{SN}}^{2}=\sum_{i=1}^{N_{SN}}\frac{(\mu
_{th}(z_{i})-\mu _{obs,i})^{2}}{\sigma _{SN, i}^{2}},
\label{eq:chiSN}
\end{equation}
where $\sigma_{SN, i}$ is the error on $\mu _{obs,i}$. The
parameter $\mu_{0}$ is a nuisance parameter which depends on the
Hubble constant. One can perform a standard marginalization on
$\mu_{0}$. Otherwise, following \cite{DiPietro+03, Nesseris+05,
Perivolaropoulos05, Wei10}, it is easy to check that the
$\chi^{2}$ in Eq. (\ref{eq:chiSN}) is equivalent to the following
function
\begin{equation}
\tilde{\chi} _{\mathrm{SN}}^{2}= \tilde{A} -
\frac{\tilde{B}^{2}}{\tilde{C}}, \label{eq:chiSN_2}
\end{equation}
where
\begin{align}
\tilde{A} & = \sum_{i=1}^{N_{SN}}\frac{(\mu _{th}(z_{i}, \mu_{0} =
0)- \mu _{obs,i})^{2}}{\sigma _{SN, i}^{2}}~, \\
\tilde{B} & = \sum_{i=1}^{N_{SN}}\frac{(\mu _{th}(z_{i}, \mu_{0} =
0)- \mu _{obs,i})}{\sigma _{SN, i}^{2}}~, \\
\tilde{C} & = \sum_{i=1}^{N_{SN}}\frac{1}{\sigma _{SN, i}^{2}}~.
\end{align}
This new function does not depend on $\mu_{0}$, allowing us to
drop the contribution from the Hubble constant.

\subsection{Baryon Acoustic Oscillation (A)}

In the large-scale clustering of galaxies, the baryon acoustic
oscillation signatures could be seen as a standard ruler providing
the other important way to constrain the expansion history of the
Universe. We use the measurement of the BAO peak from a
spectroscopic sample of 46,748 luminous red galaxies (LRGs)
observations of SDSS to test cosmology \cite{Eisenstein+05}, which
gives the value of $A=0.469(n_{s}/0.98)^{-0.35}\pm 0.017$ at
$z_{\mathrm{BAO}}=0.35$ where $n_{s}=0.96$ \cite{komatsu+09}. The
expression of $A$ can be written as
\begin{eqnarray}
A = \frac{\sqrt{\Om}}{(H(\zB)/H_{0})^{\frac{1}{3}}} \left [%
\frac{1}{\zB}\int_{0}^{\zB}\frac{dz^{^{\prime }}}{%
H(z^{^{\prime }})/H_{0}} \right ]^{\frac{2}{3}}~, \label{eq:BAO}
\end{eqnarray}
which is evidently independent on $H_{0}$, and the relative
$\chi^{2}$ function is
\begin{equation}
\chi
_{BAO}^{2}=\Big(\frac{A-0.469(n_{s}/0.98)^{-0.35}}{0.017}\Big)^{2}.
\label{chiBAO}
\end{equation}

\subsection{CMB Data: the shift parameter (R)}

The measurement of CMB anisotropies represents a powerful tool to
constrain cosmological parameters, however, using the full data of
CMB is time consuming, thus as an alternative it is common to rely
on the measurement of the shift parameter $R$. The CMB shift
parameter may provide an effective way to constrain the parameters
of DE models since it has the very large redshift distribution,
which allows to constrain the evolution of DE very well. The shift
parameter $R$ which is derived from the CMB data takes the form as
\begin{eqnarray}
R &=&\sqrt{\Om}\int_{0}^{\zCMB}\frac{dz^{\prime }}{%
H(z^{\prime })/H_{0}},  \label{eq:shift parameter}
\end{eqnarray}
where is $\zCMB = 1090$ and the observed value for Eq.
(\ref{eq:shift parameter}) has been updated to $R=1.71\pm 0.019$
from WMAP5 \cite{komatsu+09}. The $\chi^{2}$ function is
\begin{equation}
\chi _{\mathrm{CMB}}^{2}=\Big(\frac{R-1.71}{0.019}\Big)^{2}.
\label{chiCMB}
\end{equation}

\begin{figure}[t]
\vspace{0.5cm}
\includegraphics[scale=0.6,angle=0]{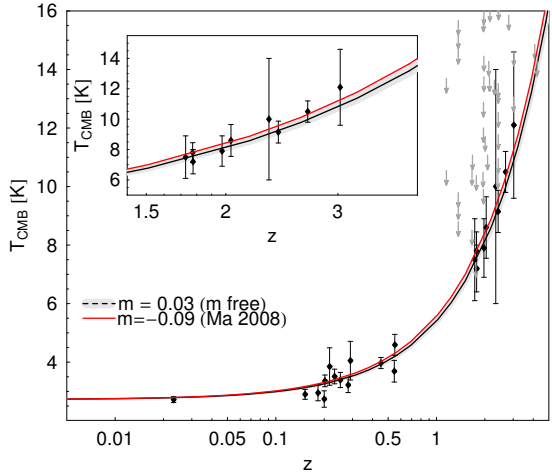}
\caption{Cosmic microwave background temperature as a function of
the redshift. The black points with bars are the full collection
of measurements from Luzzi et al. \cite{luz09} and Noterdaeme et
al. \cite{Noterdaeme+11}. The gray arrows represent the upper
limits derived from the analysis of atomic carbon (see
\cite{Noterdaeme+11} for details). The black line is the best fit
result ($m=0.03$), while the gray region is the $1\sigma$
uncertainty. The red line is the best fit recovered from Ma
\cite{ma}. The inset panel show a magnified vision of the higher
redshift region of the plot.} \label{fig:figT}
\end{figure}

\subsection{Observational $H(z)$ Data (OHD)}

By using the differential ages of passively evolving galaxies
determined from the Gemini Deep Deep Survey (GDDS) and archival
data \cite{GDDS}, Simon et al. determined H(z) in the range
$0<z<1.8$ \cite{Simon+05}. The 9 observational $H_{obs, i}$
datapoints can be obtained from \cite{Simon+05, SR06, WZ07}. The
$\chi ^{2}$ statistics for these $H(z)$ data is
\begin{equation}
\chi _{\mathrm{OHD}}^{2}=\sum_{i=1}^{9}\frac{(\log_{10}H(z_{i})-
\log_{10}H_{obs, i})^{2}}{\sigma _{OHD, i}^{2}}~, \label{eq:chiOHD}
\end{equation}
where $\sigma _{OHD, i}$ is the error on $\log_{10}H_{obs, i}$.
Following the same procedure adopted in Sec.\ref{sec:SN}, to
marginalize with respect to $H_{0}$, we replace the $\chi^{2}$ in
Eq. (\ref{eq:chiOHD}) with Eq. (\ref{eq:chiSN_2}), where
\begin{align}
\tilde{A} & = \sum_{i=1}^{9}\frac{(\log_{10}H(z_{i},H_{0}=1)-
\log_{10}H_{obs, i})^{2}}{\sigma _{OHD, i}^{2}}~, \\
\tilde{B} & = \sum_{i=1}^{9}\frac{(\log_{10}H(z_{i},H_{0}=1)-
\log_{10}H_{obs, i})}{\sigma _{OHD, i}^{2}}~, \\
\tilde{C} & = \sum_{i=1}^{9}\frac{1}{\sigma _{OHD, i}^{2}}~.
\end{align}

\section{Results}\label{sec:results}

We discuss in this section the best fitted
values for our model parameters. We first concentrate on the
temperature-redshift relation, deriving constraints on both $m$
and $\gamma$. In order to constrain the present matter density
\Om\ and $m$ we add the set of observational probes we have listed
in the previous section, which are primarily linked to the Hubble
parameter $H(z)$, and thus much more sensitive to matter
component.

\subsection{Constraints from $T(z)-z$ relation}

As a primary test, following the same line in Jetzer et al.
\cite{Jetzer+11}, we have compared the CMB temperature predicted
(see Eq. (\ref{ma3})), with the updated collection of
multi-redshift measurements of $T_{\rm CMB}$ we have discussed in
the previous section. We set $T_0 \, = \, 2.725 \ {\rm K}$, which
is quite well determined in the literature \cite{mat99}, and the
matter density $\Om=0.273$ to the value inferred in Komatsu et al.
\cite{komatsu+09}. As we have shown in Sect. \ref{sec:T_relation},
the temperature is not sensitive to changes in \Om\ in the region
where $\Om \gsim \, 0.1-0.2$ in which it is constrained to lie
from other probes. Thus, we are sure that our estimates are robust
and in this section we will not discuss the constraints on \Om\
coming from the fitting of temperature-redshift relation. If we
take $\gamma=4/3$, then we find $m = 0.03^{+0.08}_{-0.09}$, which
is lower than the estimated value of $m=0.09 \pm 0.10$ in
\cite{Jetzer+11}, but fully consistent within uncertainties, and
also pretty consistent with $m = 0$. In Fig. \ref{fig:figT} the
temperature measurements (together with some upper limits) are
shown, and our best fitted result is plotted and compared with the
$m=-0.09$ result in Ma \cite{ma}. The value we have found
corresponds to an effective equation of state $\weff = -0.99 \pm
0.03$, consistent with $\weff = -1$, and the transition redshift
is $z_{T} = 0.78 \pm 0.08$. In order to check the impact of
redshift distribution we separate the data in two redshift bins
with $z < 0.6$ (SZ data only) and $z \geq 0.6$ (QSO adsorption
lines only), finding the best fitted values
$m=0.12^{+0.12}_{-0.13}$ and $-0.05^{+0.12}_{-0.14}$,
respectively. Although the uncertainties are very high and no
statistically relevant conclusion can be reached, these results
give some indications of a mild trend with lower redshift data
preferring a DE decaying into matter and radiation, while data at
$z > 0.6$ point to an opposite behavior. These results could be
interpreted in a different way, in fact, the differences found
could be due not to the different redshift coverage, but to some
particular biases in the two kind of observations, SZ vs QSO
adsorption lines. An indication of this suggestion comes if we
divide the sample in three subsamples: 1) SZ data, 2) the data
from the analysis of the fine structure of atomic carbon (AC), and
3) the measurements based on the rotational excitation of CO
molecules in \cite{Noterdaeme+11}. If we fit the model to the
combined SZ+AC and SZ+CO samples we find $m =
0.11_{-0.11}^{+0.11}$ and $m= 0.03_{-0.10}^{+0.09}$, respectively.
Because of the larger measurement errors, the AC data affect very
little the results when only SZ measurements are adopted, and the
result for the second sample shows that SZ and CO data mainly
constrain $m$, and AC simply gives a tiny reduction on the errors.
Of course, larger data samples would be needed to provide
definitive answers.

Adopting a constant value for the ratio $\psi_{\gamma}/3n_{\gamma}H=\beta$
Lima et al. \cite{lima+00} have found the simple relation,
\begin{equation}
T(z)=T_0 (1+z)^{1-\beta}~. \label{d1}
\end{equation}
If we fit all the sample we find $\beta = -0.002 \pm 0.03$, while
for the low and high redshift subsamples we have $\beta = 0.06 \pm
0.08$ and $-0.01 \pm 0.03$, respectively. These results are
qualitatively consistent with what found in Noterdaeme et al.
\cite{Noterdaeme+11}, and points to a similar trend as the one
discussed above\footnote{Although we use the same datasample as in
\cite{Noterdaeme+11}, we find some differences (although
very minor), which could be possibly due to the way the asymmetric
errors in QSO absorption line data are accounted for. In
particular, we have adopted as error in the fit the average of the
two errors.}.

\begin{figure}[t]
\vspace{0.5cm}
\includegraphics[scale=0.6,angle=0]{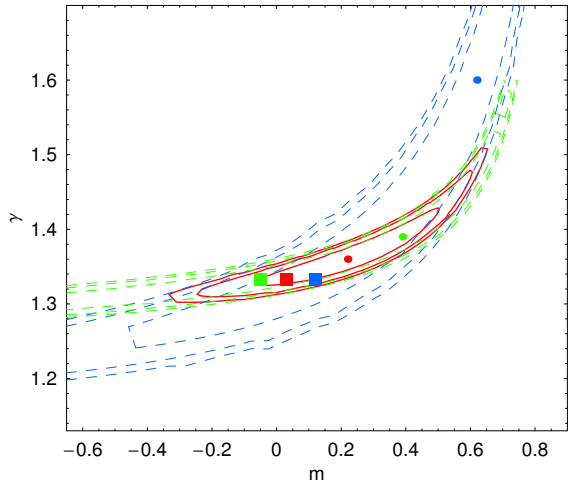}
\caption{The $68\%$, $90\%$ and $95\%$ confidence limit contours
in the $m-\gamma$ plane from the fitting of temperature
measurements. The red, blue and green contours are relative to the
fit using all the sample, data for $z < 0.6$ and the ones for
$z>0.6$, respectively. The points are the best fitted values,
while boxes are the results when $\gamma = 4/3$.} \label{fig:fig2}
\end{figure}

If $\gamma$ is left free to change, we obtain the contours shown
in Fig. \ref{fig:fig2}. In all the redshift samples, for the best
fitted value it turns out that $\gamma > 4/3$ and $m$ is systematically more positive.
Adopting the whole sample, the CL contours are broad, and from the
marginalization with respect to the second parameter we find
$\gamma = 1.35_{-0.03}^{+0.03}$ and $m= 0.25_{-0.17}^{+0.23}$,
which corresponds to an effective equation of state $\weff = -0.92
\pm 0.07$ and the transition redshift is $z_{T} = 1.1 \pm 0.6$.
When the two subsamples are adopted, wide confidence contours are
found, particularly for the $z \geq 0.6$ sample, for which the
contours at very low $m$ are not closed.  We obtain $\gamma =
1.3_{-0.1}^{+0.2}$ and $1.26_{-0.01}^{+0.01}$, while $m =
0.8_{-0.3}^{+0.1}$ and $0.6_{-1.0}^{+0.1}$, respectively for the
low and high-z samples\footnote{We notice that due to the
particular form of the CL contours the best fit quantities derived
from the maximum of the marginalized likelihood can be different
from the $\chi^{2}$ minimum. In fact, we find $\gamma = 1.6$ and
$1.4$, while $m=0.6$ and $m=0.4$ for the two subsamples, which
differ from the ML values reported in the text, but consistent
within errors, due to the large uncertainties.}.

\begin{table}[t]
\centering \caption{Maximum likelihood parameter and $1\sigma$
uncertainties of $m$ for $\Om = 0.273$ and for $m$ and \Om\ when
this last is left free to vary. The legend of the symbols is: T =
Temperature, SN = Supernovae Ia, A = Baryon Acoustic oscillation
parameter, R = Shift parameter, OHD = Observational $H(z)$
data.}\label{tab:Omfixed} \vspace{0.5cm}
\begin{tabular}{lccc} \hline \hline
 \rm Model & $\Om = 0.273$ & \multicolumn{2}{c}{\Om\ free}   \\
  \hline
  & $m$ & $\Om$ & m  \\
  \hline
SN & $0.01^{+0.16}_{-0.17}$ &  \, $0.25_{-0.09}^{+0.07}$ \, & \, $-0.2_{-0.7}^{+0.6}$ \, \\
SN+T & $0.03^{+0.07}_{-0.08}$ & $0.28_{-0.02}^{+0.03}$ & $0.03_{-0.08}^{+0.10}$ \\
SN+A & $0.01^{+0.16}_{-0.17}$ & $0.28_{-0.03}^{+0.02}$ & $0.03_{-0.24}^{+0.19}$\\
SN+A+R & $-0.04^{+0.03}_{-0.04}$ & $0.27_{-0.01}^{+0.02}$ & $-0.04_{-0.04}^{+0.04}$\\
SN+A+R+T & $-0.03^{+0.03}_{-0.03}$ & $0.27_{-0.01}^{+0.02}$ & $-0.03_{-0.03}^{+0.03}$ \\
SN+A+OHD & $-0.03^{+0.14}_{-0.16}$ & $0.27_{-0.02}^{+0.03}$ & $-0.02_{-0.18}^{+0.22}$  \\
SN+A+OHD+R & $-0.04^{+0.03}_{-0.04}$ & $0.27_{-0.01}^{+0.02}$ & $-0.04_{-0.04}^{+0.03}$ \\
 \hline \hline
\end{tabular}
\end{table}

\begin{figure*}[t]
\vspace{0.5cm}
\includegraphics[scale=0.6,angle=0]{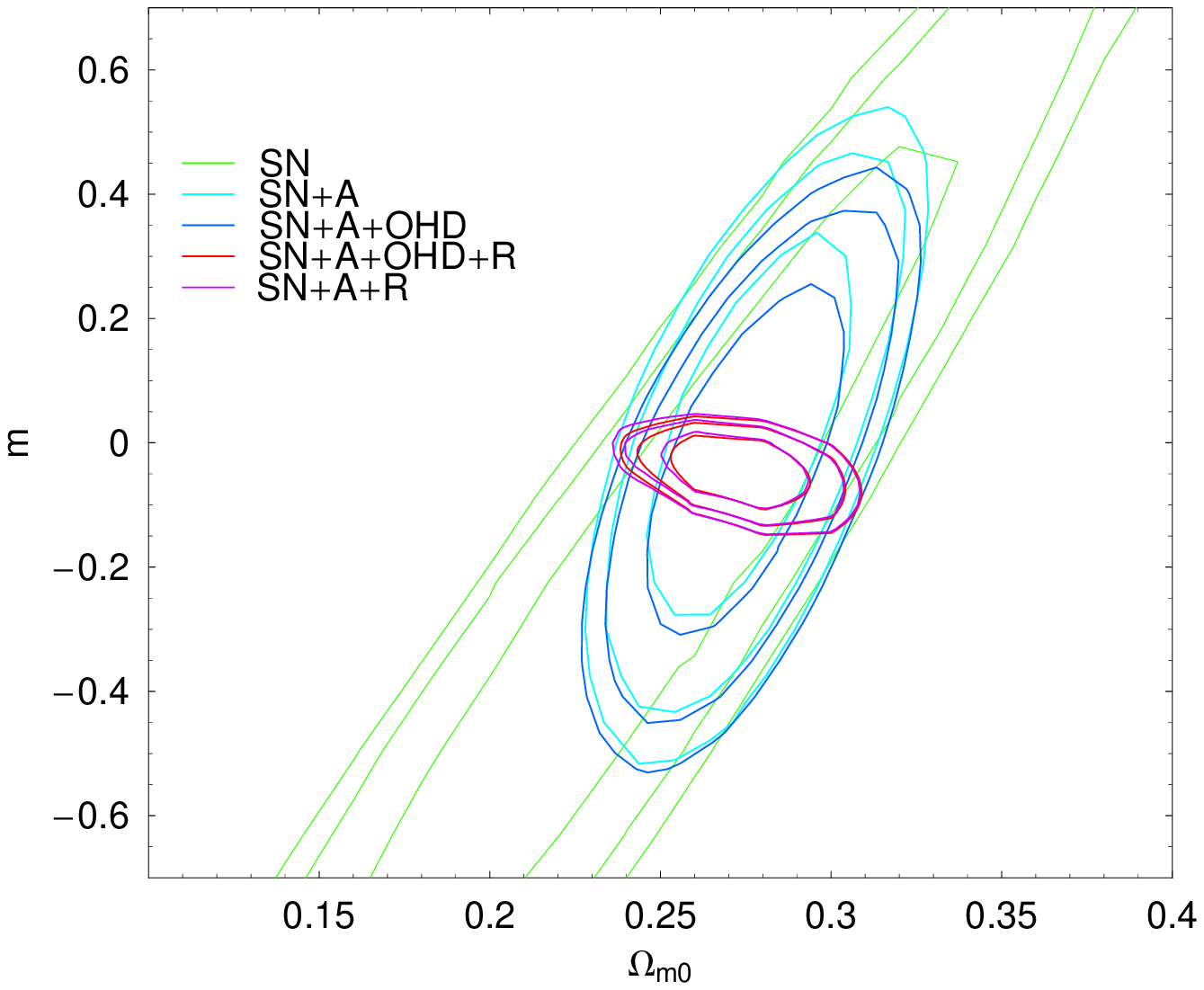}
\includegraphics[scale=0.6,angle=0]{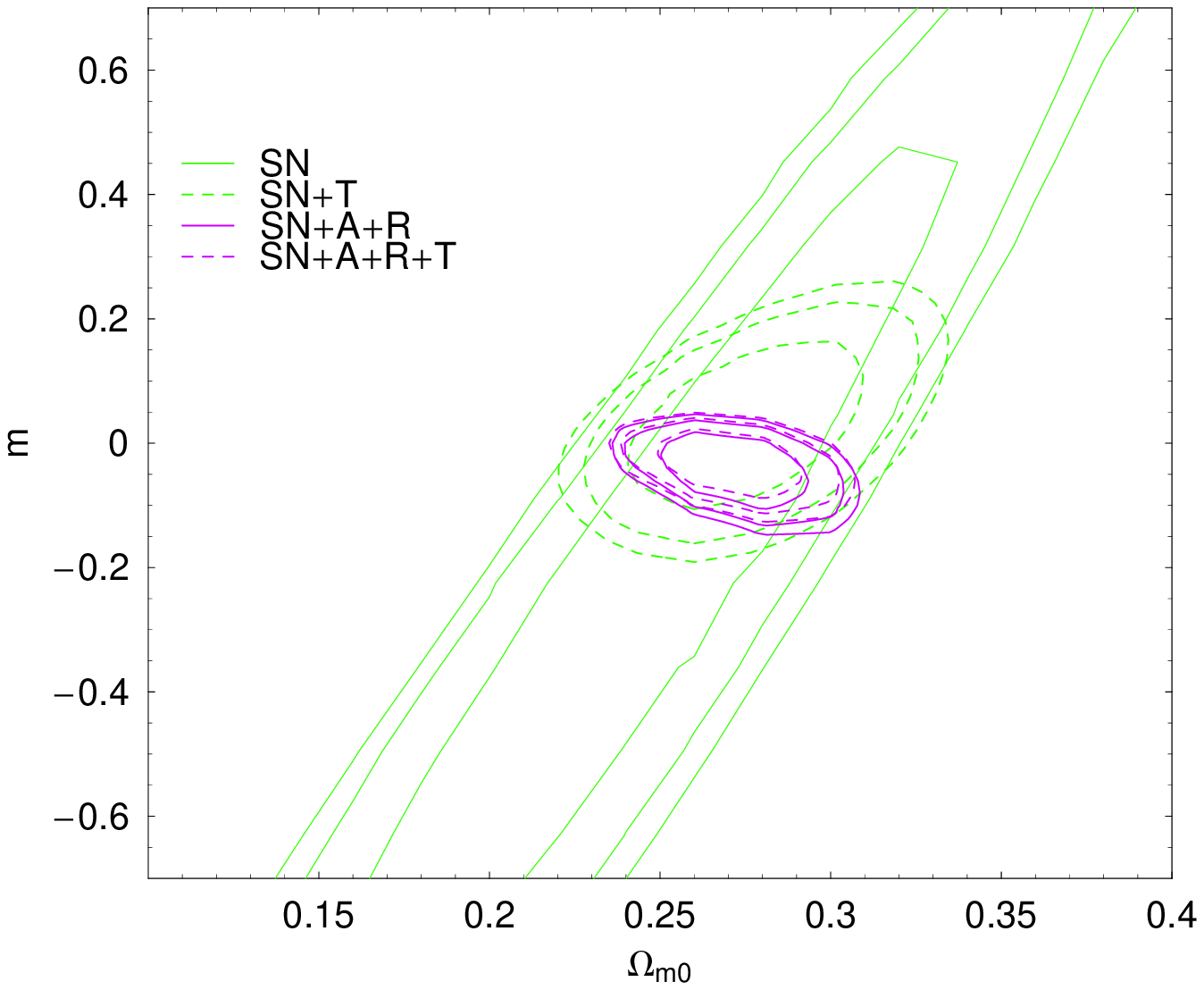}
\caption{The $68\%$, $90\%$ and $95\%$ confidence limit contours
in the $\Om-m$ plane. See the legends for the meaning of the
symbols.} \label{fig:fig3}
\end{figure*}

\subsection{Constraints from independent measurements}

If we set $\Om = 0.273$, we can give some further constraints
(shown in Table \ref{tab:Omfixed}) on $m$, using the other
observational probes listed in Sect. \ref{sec:data} together with
temperature measurements. We note that SNs alone are not able to
constrain the value of $m$, producing very high uncertainties,
also larger than the ones obtained using the temperature
measurements alone. The Baryon Acoustic Oscillation parameter does
not help, while temperature measurements will allow to reduce the
uncertainties. These probes produce slightly positive values for
$m$, while adding the Shift parameter and/or the $H(z)$ data give
negative $m \sim -0.03,-0.04$. Anyway the uncertainties remain
very large, with the only exception of the results when the Shift
parameter is used in the fitting, $m$ being in this case greatly
constrained to $m= -0.04^{+0.03}_{-0.04}$. See Table
\ref{tab:Omfixed} for details.

When $\Om$ is left free to vary, we find the best fitted values
and CL contours shown in Table \ref{tab:Omfixed} and Fig.
\ref{fig:fig3}. Remarkably, the best fitted values of \Om\ are
perfectly in agreement with the values found in independent works
\cite{Amanullah+08, komatsu+09}. Consequently, it is not
surprising that the recovered values for $m$, except for the case
when only SNs are fitted, are still quite consistent with the
estimates obtained when \Om\ is fixed (see Table
\ref{tab:Omfixed}). If we consider the fit with SN+A+OHD+R, then
our best fitted parameters correspond to an effective equation of
state $\weff = -1.01 \pm 0.01$, in the phantom regime but fully
consistent with $\weff = -1$, and the transition redshift is
$z_{T} = 0.72 \pm 0.04$.

Adopting a standard DE model with an equation of state $p_{x} = w_{x}
\rho_{x}$, and $w_{x}$ free to vary, we have performed the fit,
finding that $w_{x} \sim -1$, consistently with our results and
still pointing to a cosmological constant as the best
description for DE.

\section{Conclusions}\label{sec:conclusions}

We have presented the properties of a variable DE model, following
the approach in Jetzer et al. \cite{Jetzer+11} (see also
\cite{lima+96, LA99, lima+00, puy, ma}). The model relies on the
assumption that the DE component is not conserved and exchanges
particles with both DM and radiation components. In particular, we
show that in order to have a viable model, DE needs to decay
besides into radiation also in matter. Motivated by these
considerations, we have concentrated the analysis on the case when
only a small fraction is exchanged with radiation. We have
dedicated particular attention to the thermodynamic properties of
the model, discussing the theoretical relation between the
radiation temperature and redshift and we have matched it to the
most updated collection of CMB temperature measurements
\cite{luz09, Jetzer+11, Noterdaeme+11}, consisting of both SZ data
and high-redshift QSO adsorption line observations. First, we have
constrained the model by setting $\gamma = 4/3$ and finding $m =
0.03^{+0.08}_{-0.09}$, consistent with our previous estimate
within uncertainties, still consistent with the standard case of a
$T \propto (1+z)$, i.e. with an effective equation of state $\weff
= -1$. When $\gamma$ is left free, then $m= 0.25_{-0.17}^{+0.23}$.
Although the large contours indicate that is not possible to find
statistically relevant departures from a standard cosmological
constant, we find that a decaying DE, with an effective equation
of state $\weff > -1$ is preferred. Then, for the first time, we
test this kind of model combining both CMB temperature
measurements and different observational data sets, like
Supernovae Ia, CMB and large-scale structure data, etc. We find
that \Om\ is almost independent on the combination of data we use,
and in the best case is constrained to $\Om =
0.27_{-0.01}^{+0.02}$. Moreover, while temperature measurements
and SN data tend to furnish $m \gsim 0$, consistently with a
decaying DE model, other datasets, like CMB data prefer the
opposite situation with $m \lsim 0$, i.e. a phantom-like model
with $\weff < -1$.

Although the present data do not allow to find strong
discrepancies with a classical cosmological constant model, we
think that future surveys at high redshift could collect further
measurements for CMB temperature, which could help us to further
constrain the temperature-redshift relation. This kind of
information, together with larger and higher quality samples of
Supernovae and better CMB data or other standard candles like
Gamma ray burst (GRBs) can allow to give independent constraints
on both matter density \Om\ and effective equation of state \weff.

We notice that if DE does not decay into radiation (corresponding
to $\epsilon=\tilde{\epsilon}=0$ and thus $m$ is no longer
constrained) then the CMB temperature will scale in the standard
way. Clearly, this would imply that if DE decays, this has to be
into DM only. On the other hand a deviation  of the CMB
temperature from the standard scaling could be interpreted as DE
decaying also into radiation. In which case with Eq. (\ref{ma3})
one can determine $m$ and/or $\tilde{\epsilon}$ and thus get some
insights on the decay mode of DE into radiation. Future data on
the CMB temperature will allow to shed light on this important
issue.

\begin{acknowledgments}
C. Tortora was supported by the Swiss National Science Foundation.
\end{acknowledgments}


\begin{thebibliography}{99}

\bibitem{Jetzer+11} P. Jetzer, D. Puy, M. Signore, C. Tortora, Gen. Relativ. Gravit., {\bf 43}, 1083 (2011)

\bibitem{perl} S. Perlmutter et al., Astrophys.J. {\bf 517}, 565 (1999)

\bibitem{reiss} A.G. Reiss et al., Astron.J. {\bf 116}, 1009 (1998)

\bibitem{debernardis} P. de Bernardis et al., Nature {\bf 404}, 955 (2000)

\bibitem{spe07} D.N. Spergel et al., Astrophys. J. Suppl. {\bf 170}, 377 (2007)

\bibitem{caldwell} R. Caldwell \& M. Kamionkowski, Ann. Rev. Nucl. Part. Sci.
{\bf 59}, 397 (2009)

\bibitem{PR88}  P.J.E. Peebles \& B. Ratra, Astrophys. J. Lett. {\bf 325}, L17
(1988)

\bibitem{RP88}  B. Ratra \& P.J.E. Peebles, Phys. Rev. D {\bf 37}, 3406  (1988)

\bibitem{SS00} V. Sahni \& A.A. Starobinsky, Int. J. Mod. Phys. {\bf D9}  373 (2000)

\bibitem{Caldwell02} R.R. Caldwell, Phys. Lett. B {\bf 545}, 23 (2002)

\bibitem{Pad03} T. Padmanabhan, Phys. Rep. {\bf 380}, 235 (2003)


\bibitem{PR03} P. J. E. Peebles and B. Rathra, Rev. Mod. Phys. {\bf 75}, 559 (2003)

\bibitem{D+05} M. Demianski, E. Piedipalumbo, C. Rubano, C.
Tortora, A\&A {\bf 431} 27D (2005)

\bibitem{CTTC06} V. F. Cardone, C. Tortora, A. Troisi, and S.
Capozziello, Phys. Rev. {\bf D73}, 043508 (2006)


\bibitem{lima+96} J.A.S. Lima, Phys. Rev. D. 54, 2571 (1996)

\bibitem{LA99} J.A.S. Lima and J.S. Alcaniz, Astron. and Astrophys. {\bf 348}, 1 (1999)

\bibitem{lima+00} J.A.S. Lima, A.I. Silva and S.M. Viegas, Mon. Not. R. Astron. Soc. 312, 747 (2000)


\bibitem{puy} D. Puy, Astron. and Astrophys. {\bf 422}, 1 (2004)

\bibitem{CW90} W. Chen, Y.S. Wu, Phys. Rev. D 41 (1990) 695; W. Chen, Y.S. Wu,
Phys. Rev. D 45 (1992) 4728, Erratum.

\bibitem{ma} Y. Ma, Nucl. Phys. B {\bf 804}, 262 (2008)



\bibitem{Weinberg72} S. Weinberg, {\it Gravitation and Cosmology: Principles and Applications
of the General Theory of Relativity} (Wiley, New York, 1972!.


\bibitem{battistelli} E.S. Battistelli et al., Astrophys. J. {\bf 580}, L101 (2002)

\bibitem{luz09} G. Luzzi, M. Shimon, L. Lamagna, Y. Rephaeli, M. De Petris, A. Conte, S. De Gregori and E. Battistelli, Astrophys. J. {\bf 705}, 1122 (2009)


\bibitem{Noterdaeme+11} P. Noterdaeme, P. Petitjean, R. Srianand, C. Ledoux, and S.
L\'opez, A\&A {\bf 526}, L7 (2011)

\bibitem{mubasher} M. Jamil, E.N. Saridakis and M.R. Setare, Phys. Rev. {\bf D 81}, 023007 (2010)

\bibitem{Mather+99} J.C. Mather et al., Astrophys. J. 512, 511 (1999).

\bibitem{ge97} J. Ge, J. Bechtold and J. Black, Astrophys. J {\bf 474}, 67 (1997)

\bibitem{sri00} R. Srianand, P. Petijean and C. Ledoux, Nature {\bf 408}, 931 (2000)

\bibitem{mol02} P. Molaro, S. Levshakov, M. Dessauges-Zavadsky and S. D'Odorico, Astron. and Astrophys. {\bf 381}, L64 (2002)

\bibitem{puy93} D. Puy, G. Alecian, J. Leorat, J. Lebourlot and G. Pineau des Forets, Astron. and Astrophys. {\bf 267}, 337 (1993)

\bibitem{gal98} D. Galli and F. Palla, Astron. and Astrophys. {\bf 335}, 403 (1998)

\bibitem{sta98} P. Stancil, S. Lepp and A. Dalgarno, Astrophys. J. {\bf 509}, 1 (1998)

\bibitem{cui05} J. Cui, J. Bechtold, J. Ge and D. Meyer, Astrophys. J. {\bf 633}, 649 (2005)

\bibitem{sri08} R. Srianand, P. Noterdaeme, C. Ledoux and P. Petijean, Astron. and Astrophys. {\bf 482}, L39 (2008)



\bibitem{sun72} R. Sunyaev and Y. Zel'dovich, Comm. Ap. Sp. Phys.{\bf 4}, 173 (1972)

\bibitem{fab78} R. Fabbri, F. Melchiorri and V. Natale, Astrophys. Sp. Sci. {\bf 59}, 223 (1978)



\bibitem{Amanullah+08} R. Amanullah et al., Astron. Astrophys. {bf 486},
375 (2008)

\bibitem{DiPietro+03} E. Di Pietro, J.F. Claeskens, Mon. Not. Roy. Astron. Soc. {\bf
341}, 1299 (2003)

\bibitem{Nesseris+05}  S. Nesseris, L. Perivolaropoulos, Phys. Rev. D {\bf 72}, 123519
(2005)

\bibitem{Perivolaropoulos05} L. Perivolaropoulos, Phys. Rev. D {\bf 71}, 063503 (2005)

\bibitem{Wei10} H. Wei, PhLB, {\bf 687}, 286 (2010)

\bibitem{Eisenstein+05} D.J. Eisenstein, et al., Astrophys. J. {\bf 633} 560  (2005)

\bibitem{komatsu+09} E. Komatsu et al., Astrophys. J. Suppl. {\bf 180}, 330 (2009)

\bibitem{GDDS} R. G. Abraham et al. [GDDS Collaboration], Astron. J.
{\bf  127}, 2455 (2004)

\bibitem{Simon+05} J. Simon, L. Verde, R. Jimenez, Phys. Rev. D {\bf
71} 123001 (2005)

\bibitem{SR06} L. Samushia and B. Ratra, Astrophys. J. {\bf 650}, L5 (2006)

\bibitem{WZ07} H. Wei and S.N. Zhang, Phys. Lett. B {\bf 644}, 7 (2007)

\bibitem{mat99} J.C. Mather et al., Astrophys. J. {\bf 512}, 511 (1999)

\end{thebibliography}
\end{document}